\documentclass{aa}

\usepackage{amsmath}
\usepackage[varg]{txfonts}
\usepackage{graphicx}
\usepackage{natbib}
\usepackage{pdfpages}
\usepackage{wasysym}
\usepackage{rotating}
\bibpunct{(}{)}{;}{a}{}{,} 

\begin{document}

\title{Solar cyclic activity over the last millennium reconstructed from annual $^{14}$C data
\footnote{The reconstructed open solar flux and sunspot numbers are tabulated at the CDS via anonymous ftp
 to cdsarc.u-strasbg.fr (130.79.128.5) or via http://cdsarc.u-strasbg.fr/viz-bin/cat/J/A+A/xx/yy}}

\author{I.G. Usoskin\inst{1}
\and S.K. Solanki\inst{2,6}
\and N. Krivova\inst{2}
\and B. Hofer\inst{2}
\and G.A. Kovaltsov\inst{3}
\and L. Wacker\inst{4}
\and N. Brehm\inst{4}
\and B. Kromer\inst{5}
}

\institute{{Space Physics and Astronomy Research Unit and Sodankyl\"a Geophysical Observatory, University of Oulu, Finland}
\and {Max Planck Institute for Solar System Research, Justus-Von-Liebig-Weg 3, D-37077, G\"ottingen, Germany}
\and {Ioffe Physical-Technical Institute, St. Petersburg, Russia}
\and {Laboratory of Ion Beam Physics, ETH-Z\"urich, Z\"urich, Switzerland}
\and {Institute of Environmental Physics, Heidelberg University, Heidelberg, Germany}
\and {School of Space Research, Kyung Hee University, Yongin, Gyeonggi-Do 446 701, Republic of Korea}
}

\date{}

\abstract {}
{The 11-year solar cycle (Schwabe cycle) is the dominant pattern of solar magnetic activity reflecting the
 oscillatory dynamo mechanism in the Sun's convection zone.
Solar cycles have been directly observed since 1700, while indirect proxy data suggest their existence over a much longer period
 of time but generally without resolving individual cycles and their continuity.
Here we reconstruct individual solar cycles for the last millennium using recently obtained $^{14}$C data
 and state-of-the-art models.}
{Starting with the $^{14}$C production rate determined from the so far most precise measurements of radiocarbon content
 in tree rings, solar activity is reconstructed in three physics-based steps:
 (1) Correction of the $^{14}$C production rate for the changing geomagnetic field;
 (2) Computation of the open solar magnetic flux; and (3) Conversion into sunspot numbers outside of grand minima.
All known uncertainties, including both measurement and model uncertainties are straightforwardly
 accounted for by a Monte-Carlo method.
 }
{Cyclic solar activity is reconstructed for the period 971\,--\,1900 (85 individual cycles) along with its uncertainties.
This more than doubles the number of solar cycles known from direct solar observations.
We found that lengths and strengths of well-defined cycles outside grand minima are consistent with those obtained from
 the direct sunspot observations after 1750.
The validity of the Waldmeier rule (cycles with fast rising phase tend to be stronger) is confirmed
 at a highly significant level.
Solar activity is found to be in a deep grand minimum when the activity is
 mostly below the sunspot formation threshold, during about 250 years.
Therefore, although considerable cyclic variability in $^{14}$C is seen even during grand minima,
individual solar cycles can hardly be reliably resolved therein.
Three potential solar particle events, ca. 994, 1052 and 1279 AD, are shown to occur around the maximum phases of solar cycles.}
{A new about 1000-year long solar activity reconstruction, in the form of annual (pseudo) sunspot numbers with full
 assessment of all known uncertainties, is presented
 based on new high-precision $\Delta^{14}$C measurements and state-of-the-art models,
 more than doubling the number of individually resolved solar cycles.
This forms a solid basis for new, more detailed studies of solar variability.}

\titlerunning{1000-year sunspot series}
\maketitle

\section{Introduction}

Cyclic variability with a period of about 11 years (\textit{Schwabe} cycle) is the dominant pattern of solar
 magnetic activity \citep{hathawayLR} reflecting the oscillating dynamo mechanism in the solar convection zone \citep{charbonneauLR}.
However, the 11-year cycle is far from a perfect sine-wave and varies both in magnitude and length on a longer
 time scale \citep{usoskin_LR_17}.
Historically, the most common and the longest index of solar magnetic activity is the synthetic sunspot number (SN) based on direct
 solar observations by a cohort of astronomers worldwide since 1610 \citep{vaquero16}.
Although the SN series is somewhat uncertain before ca. 1900 \citep{clette14}, it clearly depicts the dominance of the
 Schwabe cyclicity and its variability.

The overall level of solar activity and its secular variability over the last ten millennia
 has been reconstructed from decadaly resolved cosmogenic radioisotopes $^{14}$C and $^{10}$Be
  \citep{solanki_Nat_04,usoskin_AA_04,vonmoos06,delaygue11,steinhilber12,usoskin_AA_16,wu18}, but the
  11-year cycle cannot be resolved from these datasets.
Still, the existence of the 11-year solar cycle before 1610 can be found from cosmogenic
 radioisotopes for some periods \citep[e.g.,][]{miyake_JGR_13,guettler13a,muscheler16} with special emphasis
  on the 11-year cycle during grand minima \citep{beer98,miyahara04,moriya19,fogtmann19,fogtmann20}.
The existence of climate cyclicity with an appropriate period was found also in fossil data for previous
 epochs \citep[e.g.,][]{lutahrdt17,li18}.
However, the continuity of the 11-year cyclicity was not proven, as most of those studies were based
 on spectral analyses of the data \citep[e.g.,][]{eastoe19}, showing spectral peaks
 in the range of 10\,--\,12 years but did not resolve individual solar cycles.
Therefore, until recently we knew individual solar cycles only for the last 410 years since 1610,
 including a $\approx$70-year spotless period of the Maunder minimum.

A major breakthrough has been made by \citet{brehm21} who measured annually-resolved $\Delta^{14}$C
 in tree rings for the last millennium since the mid-10th century, with unprecedented accuracy.
This dataset reveals continuous solar cycles for the last millennium, at least outside of the
 grand solar minima, and displays three abrupt enhancements
 potentially associated with solar particle events (SPEs): one known event in 994 AD \citep{miyake13} and two new
 ones in 1052 and 1279 AD.
\citet{brehm21} provided an estimate of the solar modulation potential $\phi$, which characterizes the flux intensity
 of galactic cosmic rays, but whose physical interpretation is unclear \citep{caballero04,usoskin_Phi_05,herbst10,asvestari_JGR_17}.
It is not straightforward to convert $\phi$ into quantities useful for Sun-Earth relations, such as solar magnetic flux or
 solar irradiance.

Here we provide, for the first time, a physics-based quantitative reconstruction of solar magnetic activity
 since 971 AD, at a cadence that allows individual solar cycles to be resolved.
Provided quantities are open solar magnetic flux (OSF, henceforth denoted as $F$o)
 and the sunspot number (at least for the times outside deep grand minima).
Within the grand minima we provide solar activity in the form of pseudo-sunspot numbers, as described below.
The reconstruction includes a sequence of model steps, each based on the up-to-date knowledge of
 physical processes involved and the related uncertainties.
We emphasize that the reconstruction does not involve any freely tunable ad-hoc parameters or normalization, since all
 model parameters were determined independently of this reconstruction.

\section{Data}
\label{S:data}
\begin{figure}
\centerline{\includegraphics[width=\columnwidth]{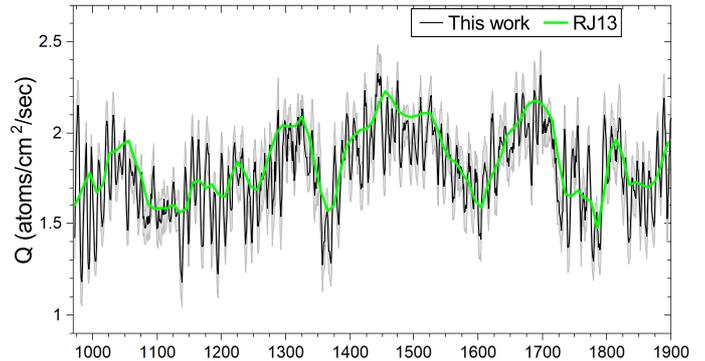}} 	
\caption{The $^{14}$C production rate $Q$ (black curve) with $1\sigma$ uncertainties, obtained from \citet{brehm21}, used here,
 after corrections for the 994, 1052 and 1279 AD events and smoothing (see Section~\ref{S:data}).
The green RJ13 curve depicts $Q$ from \citet{roth13}.
}
\label{Fig:Q}
\end{figure}

The series of the production rate of radiocarbon (denoted as $Q$ henceforth)
 data for 971\,--\,1900 with uncertainties was obtained from \citet{brehm21}
 who computed it from the measured annual $\Delta^{14}$C measurements with a pseudo-monthly
 resolution applying a most recent carbon-cycle box model \citep{buentgen18} and correcting
 for the Suess effect (dilution of atmospheric concentration of $^{14}$C because of the
 fossil fuel burning since the late 19th century).
It is consistent with the earlier decadal $^{14}$C production rate \citep{roth13} but
 resolving individual solar cycles.
This data set contains several short 1-year interpolations corresponding to gaps in the raw $\Delta^{14}$C dataset
 (1203, 1277, 1304, 1309, 1312, 1578, 1645, 1702, 1715) and one longer gap (6 years between 1043 and 1048).
The solar cycle corresponding to the latter gap is marked as unreliable.
During the analyzed period, an extreme solar event occurred ca. 994 \citep{mekhaldi15,miyake13} that may distort the $^{14}$C production.
The effect of this event was removed by subtracting the modelled production of $\Delta Q=3.9$ at/cm$^2$/sec \citep{mekhaldi15}
 spread over the years 992, 993, 994 as 1, 1.9, 1 at/cm$^2$/sec, respectively, before further analysis.
Similarly we have removed the potential events of 1052 considered as 0.65$\times$ 994 AD event, starting in 1051,
 and 1279, considered as 0.8$\times$ 994 AD event, starting in 1279 \citep[see][]{brehm21}.
However, since the exact strength of these events is not well-known yet, we mark the corresponding cycles as not
 well-defined.
The effect of removing these events is shown in Section~\ref{Sec:994}.
Because of the high level of noise in the $Q$-series, it was slightly smoothed (Savitzky-Golay filter of order 3 and framelength 9)
 before further processing.
This $Q$-series is shown in Figure~\ref{Fig:Q}.

\section{The method}
\label{Sec:method}
The method of sunspot activity reconstruction consists of three consecutive steps
\begin{equation}
Q \stackrel{(1)}{\longrightarrow} Q^* \stackrel{(2)}{\longrightarrow} F_{\rm o} \stackrel{(3)}{\longrightarrow} {\rm SN},\nonumber
\label{eq:method}
\end{equation}
each performed 10000 times in a Monte-Carlo (MC) procedure as described below.
Henceforth, the index $i$ denotes the number of the realization, and $j$ the year within the time series ($j$=1 corresponds to 971 AD).

\subsection{Step (1): Reducing $Q$ to the modern geomagnetic field}
\label{Sec:Q2Q775}

\begin{figure}
\centerline{\includegraphics[width=\columnwidth]{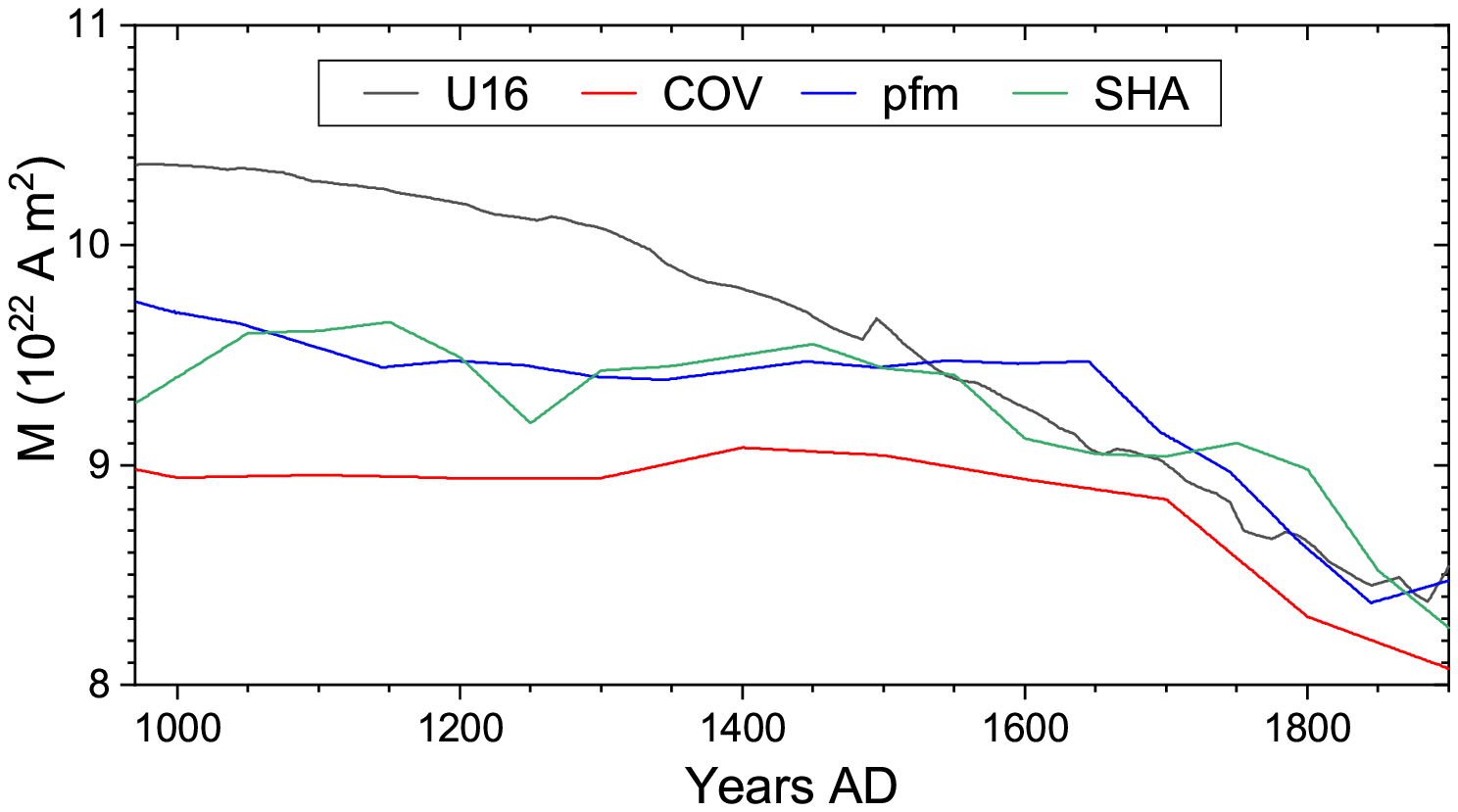}} 	
\caption{Time variability of the virtual axial dipole moment (VADM) as presented in archeomagnetic models
 considered here U16 \citep{usoskin_AA_16}, COV-ARCH \citep{hellio18},
 pfm9k.1b \citep{nilsson14} and SHA \citep{pavon14}.
}
\label{Fig:M_data}
\end{figure}

To remove the effect of the variable geomagnetic shielding, we reduced the production rate $Q$, obtained by \citet{brehm21},
 to $Q^*$ corresponding to that for the reference geomagnetic field with the fixed dipole moment $M_0$.
The following sub-steps were used:
\begin{align}
(1{\rm a}) &:& M_{i,j} = M_j^{(k_i)},\,\,\, k_i=r_{[1-4]} \nonumber\\
(1{\rm b}) &:& Q_{i,j}(M_{i,j}) \rightarrow Q^*_{i,j}(M_0)
\label{eq:step2}
\end{align}
First (step~\ref{eq:step2}a), one of four archeomagnetic models, namely U16 \citep{usoskin_AA_16}, COV \citep{hellio18},
 pfm9k.1b \citep{nilsson14} and SHA \citep{pavon14} was randomly chosen for each realization $i$.
These models (shown in Figure~\ref{Fig:M_data}) were selected as representing the diversity of the main
 groups working in paleo/archeo-magnetic recosntructions, and covering the full range of uncertainties
 of archeomagnetic models for the last millennium.
As the reference geomagnetic field we have considered recent IGRF \citep[IGRF model --][]{thebault15}
 conditions with the dipole moment $M_0$=7.75$\cdot10^{22}$ A m$^2$ corresponding to the time when
 cosmic-ray spectra were directly measured in space (see Step 3).

\begin{figure}
\centerline{\includegraphics[width=\columnwidth]{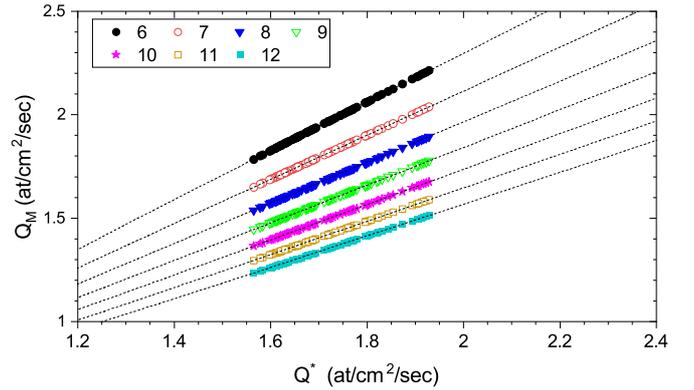}} 	
\caption{Dependence between $Q$-values computed for different geomagnetic dipole moments $M$, as denoted in the legend
 in units of $\cdot 10^{22}$ A m$^2$, and $Q^*$ at the modern-day value of $M_0$=7.75$\cdot10^{22}$ A m$^2$.
 $Q$-values were calculated using GCR spectra directly measured by AMS for May 2011 through May 2017 (see Section~\ref{Sec:AMS}).
}
\label{Fig:QM}
\end{figure}

The reduction (step~\ref{eq:step2}b) to the reference field $M_0$ was performed
 using computations based on the Galactic cosmic-ray (GCR) spectra directly measured during the last decade
  (see Section~\ref{Sec:AMS} for more details).
Using the measured spectra as an input, we computed \citep[using the production model by][]{poluianov16}
 expected values of $Q(M)$ for different values of the geomagnetic dipole moment $M$.
The relation between $Q(M)$ and $Q^*$ corresponding to the modern geomagnetic field is shown in Fig.~\ref{Fig:QM}, covering a
 modulation range of $Q$ corresponding to a full solar cycle.
One can see that the relation is nearly perfectly linear.
Thus, $Q$-values at time $t$ at which the dipole moment is $M(t)$
 can be converted into the production rate at the modern value $M_0=7.75\cdot 10^{22}$ A m$^2$, as $Q^*(t) = h\left(Q_{M}(t)\right)$,
 where $h$ is an appropriate linear relation as shown in Fig.~\ref{Fig:QM}.

\begin{figure}
\centerline{\includegraphics[width=\columnwidth]{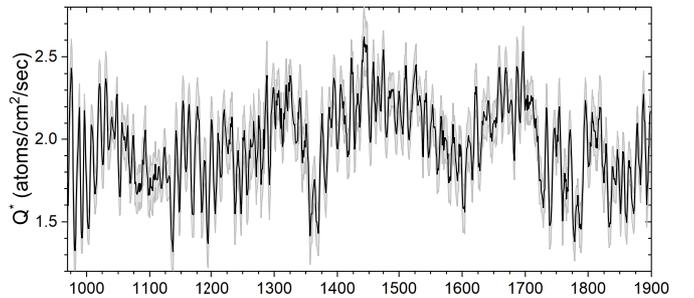}} 	
\caption{Radiocarbon production rate $Q^*$ reduced to the modern geomagnetic shielding ($M_0=7.75\, 10^{22}$ A m$^2$) with $1\sigma_{Q^*}$ errors.
}
\label{Fig:Q*}
\end{figure}

The production rate $Q^*$ reduced to the modern $M_0$ is shown in Figure~\ref{Fig:Q*}.
%

\subsection{Step (2): Open flux}
\label{Sec:OF}

This step includes conversion of the $^{14}$C production rate $Q^*$ (reduced to the reference conditions)
 to the open solar magnetic flux $F{\rm o}$:
\begin{align}
(2{\rm a})&:& F{\rm o}_{i,j} = f(Q^*_{i,j}) + R_{i,j}\cdot\sigma_{3},
\label{Eq:step3}
\end{align}
 where $f$ is a functional relating $Q^*$ to $F$o (see Equation~\ref{Eq:Fo}), $R_{i,j}$ is a normally distributed
 random number with zero mean and unity standard deviation, and $\sigma_3$ is the uncertainty of the conversion (see below).
The method used here to reconstruct the open solar flux (OSF) $F_{\rm o}$ from $Q^*$ is described below.
In contrast to earlier works, it is based on direct cosmic-ray measurements over the last decades.

\subsubsection{Use of direct space-era data}
\label{Sec:AMS}

All previous models of cosmogenic isotope production were based on theoretically modelled GCR spectra, often parameterized by the so-called
 force-field model \citep{caballero04,usoskin_Phi_05}, which however has an intrinsic uncertainty related to the local interstellar
 spectrum of GCR \citep{herbst10,asvestari_JGR_17}.
Since the spectrum of GCR beyond the Earth's atmosphere and magnetosphere was unverifiable until recently, the associated model uncertainty was present
 in all previous computations based on the force-field approximation \citep[e.g.,][]{masarik09}.
This introduced uncertainties in the level of the OSF (or other solar activity indices), sometimes leading to blind ad-hoc
 `calibrations' of the models.

The situation has been dramatically improved recently, when the space-borne \textit{Alpha Magnetic Spectrometer} experiment
 \citep[AMS -- see][]{aguilar_AMS_18} provided direct measurements of GCR energy spectra above the atmosphere over a large part of solar cycle 24 from
 May 2011 through May 2017, with a 27-day time resolution.
Not only protons, but also heavier cosmic-ray species were measured, up to iron and nickel, thus providing, for the first time, direct data on GCR
 spectra and its variability over a solar cycle.
The AMS instrument is installed onboard the International Space Station at a low orbit and spends most of the time inside the magnetosphere.
However, thanks to the inclined ($\approx 52^\circ$) orbit, it receives also low-energy cosmic rays (rigidity down to 1 GV, energy
 400 and 200 MeV/nuc for protons and heavier nuclei, respectively) over high-latitude parts of the orbit.
This makes it possible to obtain directly measured spectra of GCR ($>1$ GV) leading to the ultimate verification and, if needed,
 calibration of the $^{14}$C production model.
Contribution of even lower-energy ($<$1 GV rigidity) GCR particles to $^{14}$C production is very small
 \citep[$\approx 0.5$\%,][]{asvestari_JGR_17}, because it is limited to the small-area polar regions, while the yield function
 grows rapidly with energy.
Accordingly, we accounted for that part of the GCR spectrum not measured directly
 by AMS by extrapolating it with a best-fit force-field approximation below 1 GV.
The uncertainties related to this extrapolation are negligible ($<0.1$\%).

\begin{figure}
\centerline{\includegraphics[width=\columnwidth]{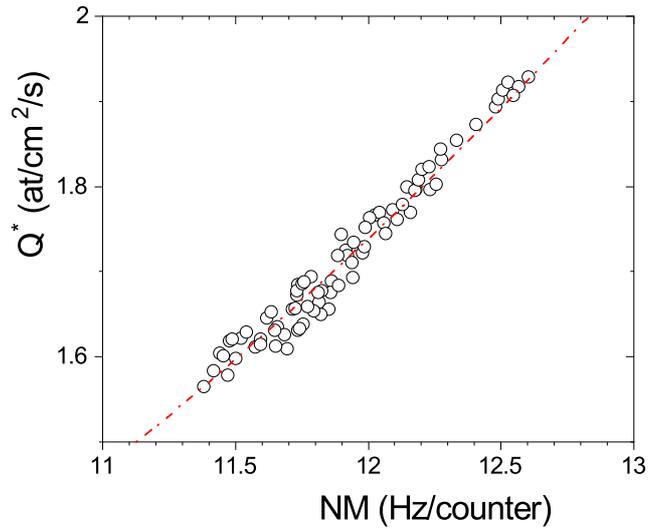}} 	
\caption{Scatter plot of the $^{14}$C global production rate $Q^*$, computed based on the AMS-02 data for the period 2011\,--\,2017,
 and a polar NM64 (Oulu) neutron monitor count rate for the same period, with the dot-dashed red line depicting the dependence (Equation~\ref{Eq:QN}).
}
\label{Fig:Q_NM}
\end{figure}

The globally averaged production rate of a cosmogenic isotope in the Earth's atmosphere at time $t$ can be calculated as
\begin{equation}
Q(t) = \sum_k{\int_0^\infty{J_k(P,t)\cdot Y^*_k(P,t)\, dP}},
\label{Eq:Q}
\end{equation}
where summation is over the type $k$ of cosmic-ray particles (protons, helium, etc),
 $J_k(P,t)$ is the rigidity (momentum over charge) spectrum of cosmic-ray particles of type $k$ near Earth
 but outside the atmosphere and magnetosphere,
%
\begin{equation}
Y^*_k(P,t) = {1\over 2}\int_{-\pi/2}^{\pi/2}{H(P_{\rm c}(\theta,t))\cdot Y_k(P)\, \cos\theta\ d\theta}
\end{equation}
 is the globally averaged yield function of the isotope production by cosmic-ray particles of type $k$ with rigidity $P$,
 $P_{\rm c}$ is the local geomagnetic cutoff rigidity \citep[e.g.,][]{elsasser56,usoskin_Geo_10} at a given geomagnetic latitude $\theta$ and time $t$,
 $H(x)$ is the Heaviside step function, and $Y_k(P)$ is the yield function of the isotope production \citep{kovaltsov12,poluianov16}.
As the global yield function $Y^*_k$ of $^{14}$C we used the one \citep{asvestari_JGR_17} based on a recent computation by \citet{poluianov16},
 energy/rigidity spectra $J_k$ of cosmic rays were taken as measured by the AMS experiment during 2011\,--\,2017,
 and the corresponding values of $Q^*$ were calculated.
Thus computed values of $Q^*$ are shown in Figure~\ref{Fig:Q_NM} against count rates of a standard polar neutron monitor (NM), viz. Oulu NM
 data record available at http://cosmicrays.oulu.fi.
Relation between them is very tight and can be parameterized as
\begin{equation}
Q^* = 0.0244\cdot N^2 - 0.2908\cdot N +1.7147,
\label{Eq:QN}
\end{equation}
where $Q^*$ is the global $^{14}$C production rate in at/cm$^2$/s for the modern epoch ($M_0$) and $N$ is the count rate of
 a polar sea-level NM in Hz/counter.

\subsubsection{Extension to 1957\,--\,2019}
\label{Sec:1957}

Using Equation~\ref{Eq:QN} and the polar NM record we have extended the expected annual $Q^*$ series backwards to 1957.
These values are plotted in Figure~\ref{Fig:Q_Fo} against OSF $F{\rm o}$ as assessed from in-situ space-borne
 data applying the kinematic correction since 1963 \citep{lockwood_2_09,owens17}, extended to recent years according to
 (Owens, personal communication, 2019) and based on geomagnetic indices \citep{lockwood_3_14}.
The values are highly significantly correlated (the Pearson's correlation coefficient $r=-0.86\pm 0.03$, $p$-value $<10^{-6}$), but the scatter is large,
 especially during the years between 1968 and 1980 (see Figure~\ref{Fig:Fo_comp}) likely because of the poor quality of in-situ solar wind data.

\subsubsection{Expected relations between $Q$ and $F$o}

\begin{figure}
\centerline{\includegraphics[width=\columnwidth]{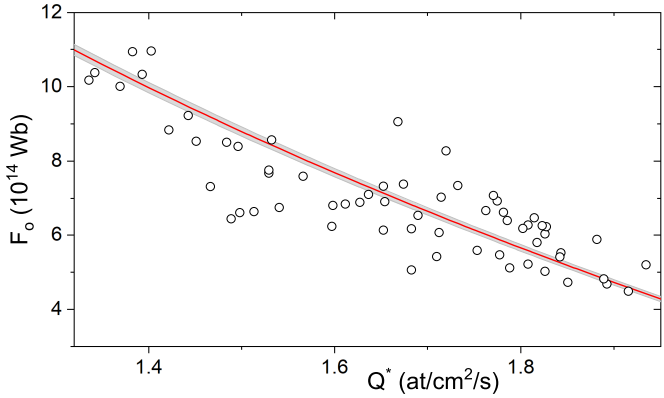}} 	
\caption{Scatter plot of the annual solar open magnetic flux $F{\rm o}$ \citep{lockwood_2_09,owens17} against the $^{14}$C production
 rate $Q^*$ estimated from NMs (see Figure~\ref{Fig:Q_NM}).
 The red curve depicts the best-fit dependence (Equation~\ref{Eq:Fo}).
}
\label{Fig:Q_Fo}
\end{figure}

Based on basic physical principles, a dependence between the isotope production rate and OSF is expected to be
 nearly exponential, $Q^* = Q_0\cdot \exp{(-F{\rm o}/\alpha)}$, which leads to
\begin{equation}
F{\rm o} = -{\alpha}\cdot \ln{\left(Q^*\over Q_0\right)},
\label{Eq:Fo}
\end{equation}
where $Q_0=2.5$ at/cm$^2$/s is the production rate in the absence of solar modulation \citep{poluianov16},
 viz. by the local interstellar spectrum ($F$o=0).
The value of $\alpha=(17.2\pm 0.2)\cdot 10^{14}$ Wb was found as the least-squares best fit to the data points
 shown in Figure~\ref{Fig:Q_Fo}.

\begin{figure}
\centerline{\includegraphics[width=\columnwidth]{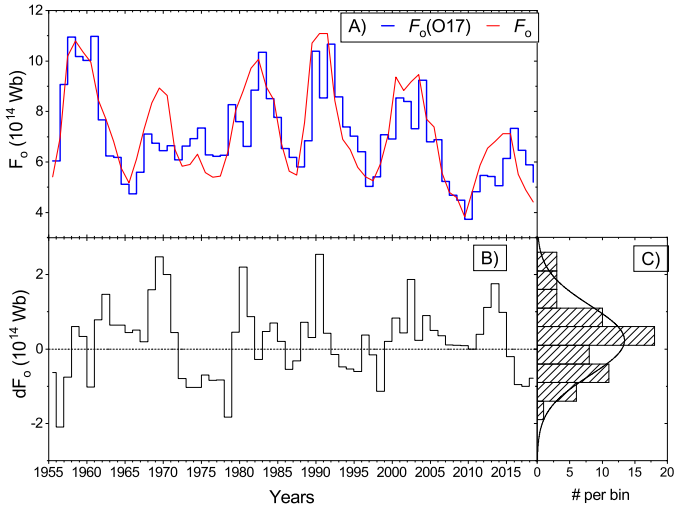}} 	
\caption{Panel A: Evolution of annual open solar flux: $F{\rm o}$(O17) based on in-situ measurements \citep{owens17} and $F{\rm o}^*$
 reconstructed here based on relations~\ref{Eq:QN} and \ref{Eq:Fo};
Panel B: difference between them $dF = F{\rm o}-F{\rm o}$(O17).
Panel C: histogram of the occurrence of $dF$ values with the best-fit Gaussian (mean $0.1\cdot10^{14}$ Wb and $\sigma=0.9\cdot 10^{14}$ Wb).
}
\label{Fig:Fo_comp}
\end{figure}

Figure~\ref{Fig:Fo_comp} shows a comparison between the OSF, derived from space-borne measurements $F{\rm o}$(O17)
 \citep{lockwood_2_09,owens17} and $F{\rm o}$ calculated here from NM data using Equations~\ref{Eq:QN} and \ref{Eq:Fo},
  as well as the difference between them (panel B).
One can see that the cycles are reproduced quite well, both in the mean level and in the amplitude.
The two OSF series agree reasonably well, with the mean difference being 0.1 and the standard deviation 0.9 (both in units of $10^{14}$ Wb).
These deviations directly enter the uncertainty of the $F{\rm o}$ reconstruction from $Q$, and
 we considered $\sigma_{3}=0.9\cdot 10^{14}$ Wb in Equation~\ref{Eq:step3}.
It is very important that the low level of the current cycle 24 (2010\,--\,2018) is reproduced correctly, suggesting that
 the secular variability is also captured by the model.

\subsubsection{Reconstruction of $F$o.}

\begin{figure}
\centerline{\includegraphics[width=\columnwidth]{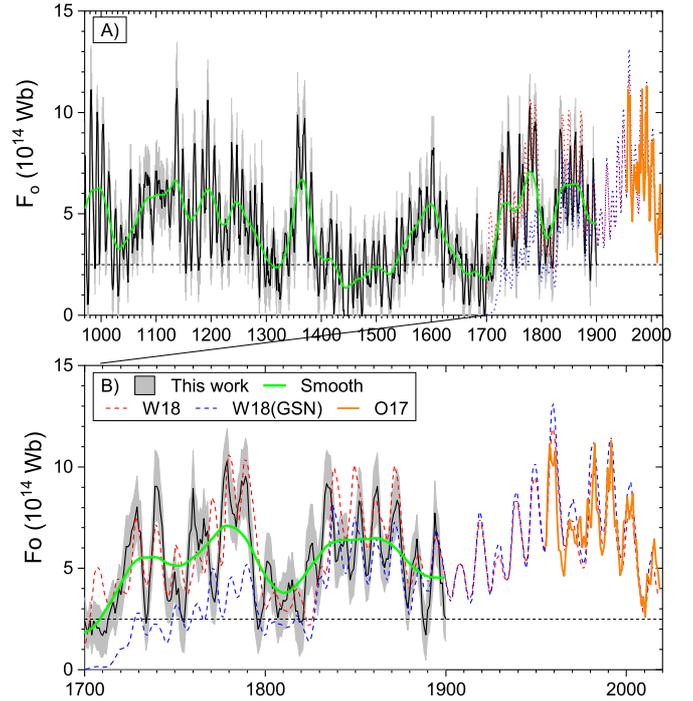}} 	
\caption{Evolution of the reconstructed annual $F_{\rm o}$.
The mean curve (black) and $1\sigma$ uncertainties (grey shaded area) were computed by 10000 Monte-Carlo realizations (see text).
The green curve represents the smoothed (22-yr SSA) variability.
The mean level of $2.5\cdot 10^{14}$ Wb defines the grand minima when solar magnetic activity drops below the sunspot formation threshold.
Other reconstructions are shown for comparison: \citep[][ -- W18]{wu_TSI_18} based on SATIRE-T model applied to ISN(v.2); the same
 SATIRE-T model but applied to the GSN (W18(GSN)), as well as the OSF reconstructed from space-based measurements \citep[][-- O17]{owens17}.
Panel B shows a zoom to the period after 1700.
The data is available in the digital tabular form in CDS tables.
}
\label{Fig:Fo}
\end{figure}

In the next step, we apply the relation (\ref{Eq:step3}) to the $Q^*$ over the whole time series, keeping in mind the found uncertainty of
 $\sigma_{F{\rm o}}= 0.9\cdot 10^{14}$ Wb and a possible systematic bias of $0.1\cdot 10^{14}$ Wb.
OSF $F{\rm o}$ reconstructed in this way is shown in Figure~\ref{Fig:Fo} with panel A depicting the whole time series and panel B
 displaying a blow-up of the period since 1700.
For comparison, several other OSF reconstructions are shown, including the $F{\rm o}$ determined from space-based measurements for
 the last decades as discussed in Section~\ref{Sec:AMS}, and two reconstructions \citep{wu_TSI_18} using the method entering the SATIRE-T model
 \citep{vieira10,krivova10}, but based on two different sunspot series: \citep[ISN(v.2) available at SILSO,][]{clette16},
 and GSN \citep{hoyt98}.
We note that these two sunspot series serve as the conservative upper and lower bounds for the uncertainties of different sunspot series \citep{usoskin_LR_17}.
Before ca. 1880s the reconstruction lies mainly between the two colored curves, being closer to the ISN-based one during the
 18th century.
Note, however, that the model  underlying the red and blue dashed curves in Figure~\ref{Fig:Fo} provides too low $F$o during extended periods
 of particularly low activity, such as grand minima, because of the limitation of the earlier OSF models.
%

\subsection{Step (3): Conversion of $F{\rm o}$ into sunspot number}
\label{Sec:step4}
The OSF $F{\rm o}$ can be estimated
 from the sunspot number using a semi-empirical model \citep[e.g.,][]{solanki02,vieira10}, which also enters
 the SATIRE-T model \citep{krivova10,wu_TSI_18} used for solar irradiance reconstruction.
It is based on solving a set of linear differential equations with several sources,
 considering a source term describing the emergence of active (ARs) and ephemeral regions (ERs) at the solar surface and their decay.
The latter includes the transfer of flux from ARs and ERs into slowly and rapidly evolving components of the OSF.
This model has recently been extended, improved and updated to take into account more recent observations of the number distribution of magnetic
 features with different levels of magnetic flux \citep{krivova21}.
Instead of just ERs it also includes the flux emerging in the form of internetwork fields, combining the two under the term
 small-scale emergences (SSEs).
All emerging magnetic bipoles are described by a single power-law distribution, which allows for a non-zero emergent flux
 even when there are no sunspots.
Thus, in contrast to the earlier SATIRE-T model, the new model returns a non-zero $F$o during grand minima, consistent with observational findings
 that it was of the order of 2$\cdot 10^{14}$ Wb and varied cyclicly during the MM
 \citep{beer98,owens12,asvestari_MNRAS_17}.

However, the inversion of the model (viz. $F{\rm o}\rightarrow\,\,$SN) is not possible analytically.
We have therefore taken the following alternative path, using a semi-empirical approach based on a statistical inversion
 of the forward model.

One shortcoming of the employed forward model remains that it relies on sunspots to determine the strength of OSF.
Therefore, at least the forward model cannot handle well fluctuations in the level of solar activity if this occurs at levels
 too low to produce sunspots.
The clear variation in the production of cosmogenic isotopes (both $^{10}$Be and $^{14}$C) during grand minima
 \citep{beer98,asvestari_MNRAS_17,brehm21}
 suggest, however, that variations in solar activity do take places at such times which are reflected in the number of sunspots
 only to a small extent \citep{vaquero15}.
In the context of the inverted model used here, such variations formally translate into variations in sunspot number, although there
 were in reality hardly any sunspots on the solar disc at that time \citep{usoskin_MM_15,vaquero15}.
To counter this, we have introduced a threshold of $F_b=2.5\cdot 10^{14}$ Wb with corresponds to zero sunspot (see Equation~\ref{EQ:Fo2SN}).
We consider that solar cycles with $F{\rm o}<2.5\cdot 10^{14}$ Wb in the smoothed
 OSF series correspond to grand minima and cannot be robustly defined.
This means that at times when the green curve in Figure~\ref{Fig:Fo} drops below the horizontal dashed line,
 we consider the variation in solar activity reconstructed
 by the model not to be dominantly caused by sunspots.
Rather, such fluctuations are then considered to be mainly due to  changes in the number of non-spot magnetic features on the solar
 surface, such as small-scale magnetic elements \citep[e.g.,][]{solanki93} forming network and plage.
Such activity cycles dominated by non-spot variations have been marked by a dashed line in Figure~\ref{Fig:SN_panels} and by italic font
 in Table~\ref{Tab:cycles} listing all the cycles.

\subsubsection{Statistical inversion}
\begin{figure}
\centerline{\includegraphics[width=\columnwidth]{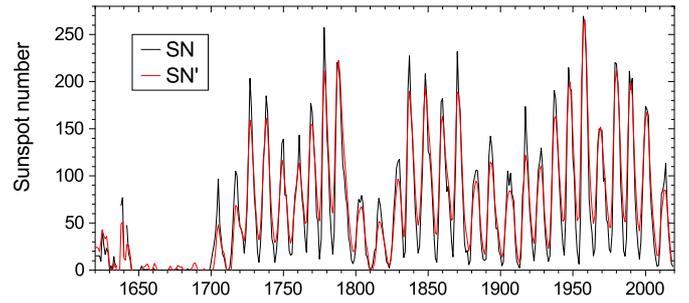}} 	
\caption{Synthetic sunspot number series, composed of ISN(v.2) after 1700 and GSN before that (black curve), and
 its `reconstruction' after the chain SN$\longrightarrow\, F$o $\longrightarrow$ SN' (red curve).}
\label{Fig:SN_test}
\end{figure}

First, we composed a synthetic series of annual sunspot numbers (SN) based on the ISN(v.2) \citep{clette16} since 1700 and
 a scaled GSN \citep{hoyt98} for the period 1610\,--\,1699.
The latter is needed, since the ISN does not cover the Maunder minimum, and we want to include this low-activity period.
This series is shown in Figure~\ref{Fig:SN_test} as the thin black curve.

This series was split into 36 individual cycles between consecutive minima of 13-month smoothed SN
 and about 11-yr intervals during the Maunder minimum.
We produced 1000 synthetic SN-series, formed by randomly permuting these 36 solar cycles
 and computing the corresponding 1000 OSF series using the equations applied in the SATIRE-T model.
Thus, we have 1000 sets of the annual `input' SN series and the corresponding `output' OSF series.

Next, we searched for a relation which inverts the input and the output, viz. $F$o $\rightarrow$ SN using an empirical approach.
Because of the presence of distinct slow secular \citep{lockwood99}, dependent on the previous history \citep[see][]{solanki00}, and
 oscillating 11-year components in the OSF, we first decomposed each $F{\rm o}$ synthetic series into slow, $F_{\rm s}$,
 and oscillating fast $F_{\rm f}=F{\rm o}-F_{\rm s}$ components.
SN of the year $j$ reconstructed from $F{\rm o}$ is calculated as
\begin{align}
{\rm SN}_j &= a\cdot \left(F'_{{\rm s}_{j+1}}\right)^2 + b\cdot \left({F'_{{\rm f}_j}+F'_{{\rm f}_{j+1}}}\right),\nonumber \\
F' &= F{\rm o} - F_{\rm b},
\label{EQ:Fo2SN}
\end{align}
where $F'_{\rm s}$ and $F'_{\rm f}$ are the slow and fast components (both expressed in units of $10^{14}$ Wb),
 obtained as the first component of the singular spectral analysis (SSA, cutoff period 5 years) of the $F'$ series and the residual, respectively,
 and parameters are $a=2.4\pm 0.05$, $b=11.5\pm 0.1$ and $F_{\rm b}=2.2\cdot10^{14}$ Wb.
The mean rms over the 1000 series was found to be 7.8 in SN units for this set of parameters.
The use of other filters does not notably alter the final result but leads to larger error bars (rms 10\,--\,13).
Thus, the model uncertainty of this step conversion was set as $\sigma_{\rm 4}$=8.

\subsubsection{Testing the inversion}

An example of the inversion for the reference series is shown as the red curve in Figure~\ref{Fig:SN_test}.
One can see that all cycles are correctly reproduced in shape and their overall level (the mean difference is 1.9) but
 with a slightly reduced amplitude (maxima are lower, minima higher), the Pearson's correlation coefficient is 0.963, rms=15.
The Maunder minimum is reproduced very well.

\begin{figure}
\centerline{\includegraphics[width=\columnwidth]{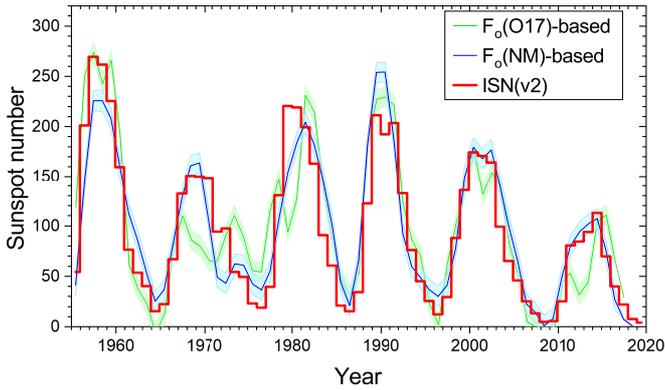}} 	
\caption{Sunspot numbers, computed using Equation~\ref{EQ:Fo2SN} from the $F_o$ data for the instrumental era
 (see Figure~\ref{Fig:Fo_comp}) reconstructed by \citep[][-- O17]{owens17} and from NM data, along with the ISN(v.2).}
\label{Fig:SN_comp}
\end{figure}

We have tested the inversion method (Eq.~\ref{EQ:Fo2SN}) using the OSF reconstruction based on in-situ data
 by \citet{owens17} (see the blue curve $F_o$(O17) in Figure~\ref{Fig:Fo_comp}).
The obtained sunspot numbers are shown in Figure~\ref{Fig:SN_comp} along with the ISN(v.2.0).
One can see that the agreement is good (mean difference $d=0.3$, Pearson's correlation coefficient $r=0.83_{-0.04}^{+0.025}$,
 rms = 35), except for the period 1963\,--\,1982, which was characterized by earlier quite uncertain magnetic-field in-situ data,
 so that they disagree with the cosmic-ray data (Figure~\ref{Fig:Fo_comp}).
When the OSF computed from the cosmic-ray data (Section~\ref{Sec:AMS}) is used, the agrement between the `reconstructed' (blue curve)
 and the actual sunspot numbers improves significantly ($d=-0.03$, $r=0.93_{-0.02}^{+0.01}$, rms = 26).
Thus, we can conclude that the method provides a good way to reconstruct the sunspot number from OSF data.
%

\subsection{Solar-activity reconstruction}
\label{S:SN_fin}

The sunspot number reconstruction was performed with 10000 random realizations, each going through steps 1\,--\,3
 and applying independent random numbers as described above in formulas \ref{eq:step2} and \ref{Eq:step3}.
The sunspot number (in the units of ISN v.2), at least for times when solar activity is characterized by a sunspot cycle,
 as during recent decades, is computed as
\begin{align}
(3{\rm a})&:& {\rm SN}_{i,j} = g(F{\rm o}_{i,j}) + R_{i,j}\cdot\sigma_{{\rm 4}},
\label{Eq:step4}
\end{align}
where the functional $g$ is defined by Equation~\ref{EQ:Fo2SN}.

\begin{figure}[t!]
\centerline{\includegraphics[width=\columnwidth]{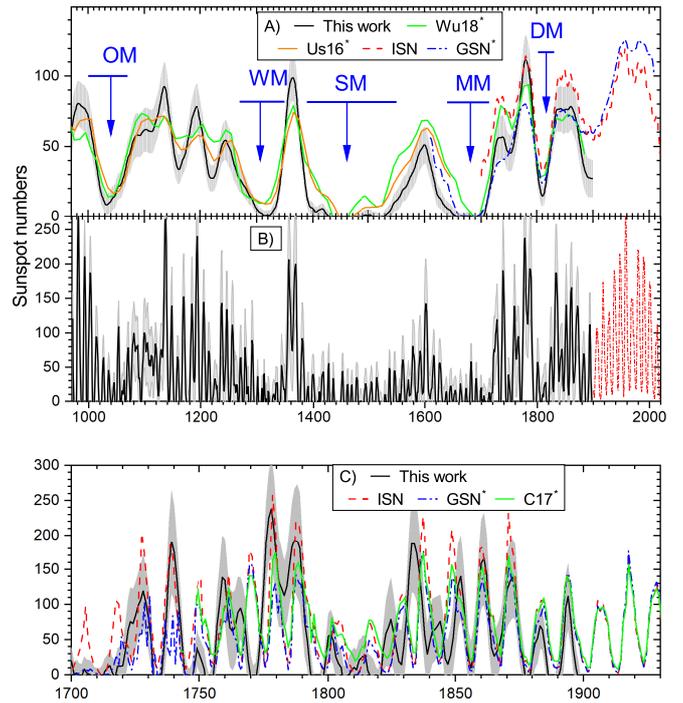}} 	
\caption{Time evolution of the reconstructed sunspot numbers (black curve with $\pm 1\sigma$ grey-shaded uncertainties)
 in comparison with other direct and indirect sunspot number series.
The final annual reconstruction is shown in panel B, and its zoom for the period after 1700 is shown in panel C.
Panel A shows the smoothed (15-yr first SSA component) annual series in comparison with other similarly smoothed or decadal series.
These series are: ISN(v2, SILSO), GSN$^*$ \citep{hoyt98}; C17$^*$ \citep{chatzistergos17}; as well as two recent
 reconstructions based on $^{14}$C -- U16$^*$ \citep{usoskin_AA_16} and W18$^*$ \citep{wu18}
(symbol $^*$ indicates that the series is scaled up by a factor 1.667 to match the ISN v.2 definition).
Blue arrows denote grand minima of solar activity: Oort (OM), Wolf (WM), Sp\"orer(SM), Maunder (MM), and Dalton (DM) minima.
The data are available in digital tabular form in CDS.
}
\label{Fig:SN}
\end{figure}

\begin{figure}[t!]
\centerline{\includegraphics[width=\columnwidth]{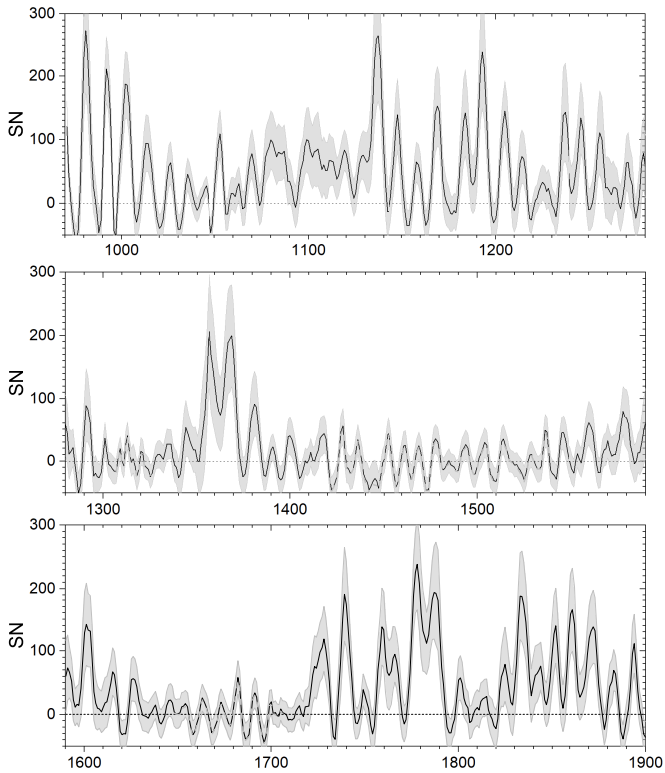}} 	
\caption{The same as Figure~\ref{Fig:SN} but split into three subsequent time intervals 310-year each.
Cycles, which are not well defined during the grand minima (see Section~\ref{Sec:step4}), are indicated with the dashed line.
}
\label{Fig:SN_panels}
\end{figure}

From 10000 thus obtained SN series we computed the mean series $\langle{\rm SN}(t)\rangle$ and its standard deviation
 $\sigma_{\rm SN}(t)$, which are considered as the final mean reconstruction and its $1\sigma$ uncertainties.
The final SN series is provided as a table at the CDS\footnote{Centre de Donn\'ees astronomiques de Strasbourg, http://cdsweb.u-strasbg.fr/about}
 and shown in Figures~\ref{Fig:SN}B and \ref{Fig:SN_panels} with $1\sigma$
 uncertainties and in comparison with other direct or indirect series.
It is gratifying that individual solar cycles are clearly resolved outside of the grand minima.
Figure~\ref{Fig:SN}A depicts a smoothed (15-yr first SSA component) SN-series with the corresponding uncertainties, in
 comparison with similarly smoothed ISN and GSN series as well as decadal $^{14}$C INTCAL-based reconstructions
  by \citet{usoskin_AA_16} and a multi-proxy based one \citet{wu18} (all series were properly scaled to match the ISN v.2 scale).
The agreement with previous cosmogenic-proxy reconstructions, including U16 \citep{usoskin_AA_16}
 and W18 \citep{wu18} ones, is very good (see Section~\ref{Sec:comp}).


%

Formally negative mean SN-values appear for about 250 years, about 80 of them are negative beyond the $1\sigma$-uncertainty
 and only 6 remains negative at the $2\sigma$ level (Figure~\ref{Fig:SN_panels}).
Thus, formally negative SN-values are statistically consistent with zeros.
For further analysis we keep the negative values since replacing them with zeros would distort the overall level.

We note that, while the uncertainties are large (mean 68\,\% uncertainty is about 34 in SN units), they are largely systematic
 (related to the model uncertainties), and thus do not affect the temporal evolution but only the level of activity.
An important advantage of the employed data is that their quality (and thus the SN reconstruction) is stable during the entire period.

\subsubsection{The effect of the 994, 1052 and 1279 AD events}
\label{Sec:994}

\begin{figure}
\centerline{\includegraphics[width=\columnwidth]{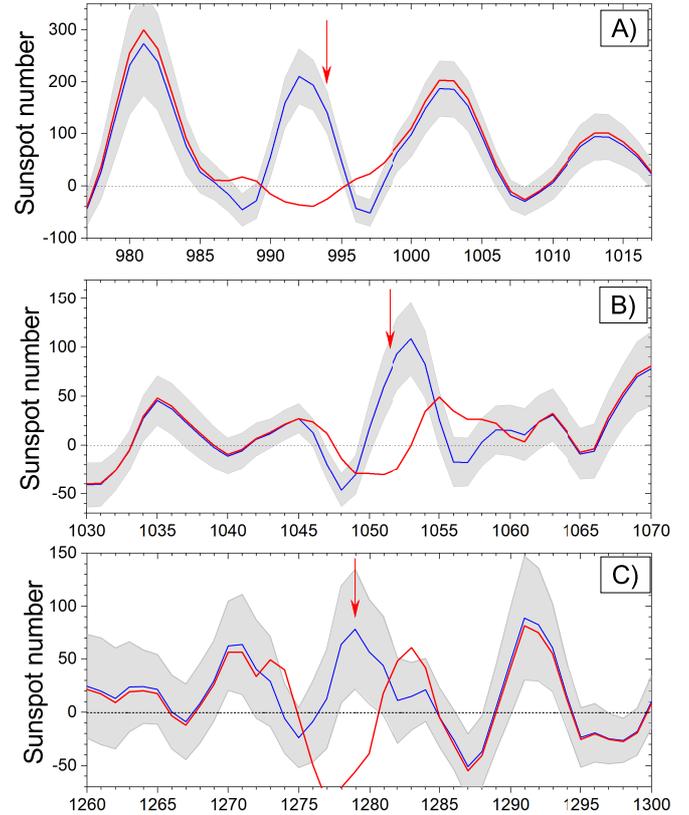}}
\caption{Evolution of the reconstructed sunspot numbers (blue curve with $\pm 1\sigma$ grey-shaded uncertainties)
 for the periods around the corrected events: 994 AD (panel A), 1052 AD (panel B) and 1279 (panel C).
The red curve depicts sunspot numbers if no correction is applied.
Red arrows indicate the time of the events.
}
\label{Fig:SN_994}
\end{figure}

One confirmed peak of additional $^{14}$C production in 994 AD \citep{miyake13} and possible peaks
 in 1052 and 1279 AD \citep{brehm21} are known during the analyzed period.
Since they are likely of non GCR origin and thus can distort the cyclic evolution of the reconstructed solar activity,
 we have removed them from the original $Q$ series as described in Section~\ref{S:data}.
The effect of the removal of the 994 AD event is shown in Figure~\ref{Fig:SN_994}A.
If the event is not corrected, a full solar cycle is `swallowed' (a bump in the $^{14}$C production
 is interpreted as very weak solar activity), while a formal (no ad-hoc tuning) correction of the effect restores a
 nearly perfect cycle with the maximum in 994 AD.
Thus, the event of 994 is found to take place at the early declining phase of a strong solar cycle.

The effect of the removal of the potential event in 1052 is shown in Figure~\ref{Fig:SN_994}B.
The correction fully restores the cyclic shape otherwise `swallowed' by the event.
The restored cycle has its maximum in 1053, thus the 1052 event took place near the maximum
 of a moderate cycle during the Oort grand minimum.

The effect of the removal of the event of 1279 is shown in Figure~\ref{Fig:SN_994}C, but it also affects the reconstructed OSF.
Again, this correction restores the cyclic shape.
As can be seen from Figure~\ref{Fig:SN_994}C, this event likely occurred around the maximum phase of a moderate cycle.

Concluding, the new annual dataset makes it possible to correctly reconstruct the overall level and phases of individual cycles
 of solar activity.
Because of uncertainties related to the event removal, we mark the related cycles as non highly reliable.
The reconstructed SN cycles are analyzed in Section~\ref{Sec:cycles}.

\section{Solar activity cycles}
\label{Sec:cycles}

We have analyzed the annual solar-activity series since 971 AD and identified individual solar cycles as presented in Table~\ref{Tab:cycles}.
A quality flag $q$ was ascribed to each cycle so that it takes values from 0 to 5 as follows:
0 -- cycle cannot be reliably identified (29 such cycles were found);
1 -- cycle is greatly distorted, at least one of its ends cannot be defined (7 cycles);
2 -- cycle can be approximately identified but either its shape or level is distorted (14 cycles);
3 -- reasonably defined cycle (10 cycles);
4 -- well-defined cycle with somewhat unclear amplitude  (19 cycle);
5 -- clear cycle in both shape and amplitude (6 cycles).

The series contains 85 full cycles (971\,--\,1900) with a mean cycle length of 10.8 years (see Section~\ref{Sec:length}).

\begin{table*}
\caption{Solar activity cycles as reconstructed here from annually resolved $^{14}$C.
Columns are: internal cycle number $n$; years of minimum $Y_{\rm min}$ and maximum $Y_{\rm max}$ of each cycle, and the corresponding
 cycle-averaged sunspot number $\langle{\rm SN}\rangle$ with $1\sigma$ uncertainties; cycle length (min-to-min) $T$ in years;
 quality flag $q$, and comments.
 Periods of deep grand minima, when solar cyclic activity drops below the sunspot formation level, are marked in \textit{italic}.
Digital version of this Table is available in CDS.
}
\begin{tabular}{cccrrrl|cccrrrl}
\hline
$n$ &  $Y_{\rm min}$ & $Y_{\rm max}$ & $\langle{\rm SN}\rangle$ & $T$ & $q$ &  Comments &
 $n$ &  $Y_{\rm min}$ & $Y_{\rm max}$ & $\langle{\rm SN}\rangle$ & $T$ & $q$ &  Comments\\
\hline
1 & 976 & 981 & 88$\pm$59 & 12 & 4 &   &  \textit{44 } & \textit{ 1457 } & \textit{ 1468 } & \textit{  -5 $\pm$ 22  } & \textit{ 8 } & \textit{ 0 } & \textit{ Sp\"orer minimum}\\
2 & 988 & 992 & 77$\pm$39 & 9 & 3 & 994 AD event  &  \textit{45 } & \textit{ 1465 } & \textit{ 1469 } & \textit{  -8 $\pm$ 21  } & \textit{ 9 } & \textit{ 0 } & \textit{ Sp\"orer minimum}\\
3 & 997 & 1002 & 82$\pm$40 & 11 & 5 &   &  \textit{46 } & \textit{ 1474 } & \textit{ 1478 } & \textit{  -0 $\pm$ 20  } & \textit{ 15 } & \textit{ 0 } & \textit{ Sp\"orer minimum}\\
4 & 1008 & 1013 & 33$\pm$33 & 12 & 5 &   &  \textit{47 } & \textit{ 1489 } & \textit{ 1494 } & \textit{  3 $\pm$ 20  } & \textit{ 10 } & \textit{ 0 } & \textit{ Sp\"orer minimum}\\
5 & 1020 & 1026 & 10$\pm$26 & 10 & 5 &   &  \textit{48 } & \textit{ 1499 } & \textit{ 1504 } & \textit{  -1 $\pm$ 22  } & \textit{ 11 } & \textit{ 0 } & \textit{ Sp\"orer minimum}\\
6 & 1030 & 1035 & 3$\pm$23 & 10 & 3 &   &  \textit{49 } & \textit{ 1510 } & \textit{ 1514 } & \textit{  0 $\pm$ 20  } & \textit{ 15 } & \textit{ 0 } & \textit{ Sp\"orer minimum}\\
7 & 1040 & 1045 & 5$\pm$16 & 8 & 1 &  1043\,--\,1048 gap &  \textit{50 } & \textit{ 1525 } & \textit{ 1529 } & \textit{  -6 $\pm$ 20  } & \textit{ 8 } & \textit{ 0 } & \textit{ Sp\"orer minimum}\\
8 & 1048 & 1053 & 33$\pm$29 & 9 & 2 &  1052 AD event & 51 & 1533 & 1536 & 5$\pm$23 & 9 & 1 &  \\
9 & 1057 & 1063 & 12$\pm$26 & 8 & 2 &   & 52 & 1542 & 1545 & 11$\pm$25 & 10 & 2 &  \\
10 & 1065 & 1070 & 35$\pm$33 & 9 & 4 &   & 53 & 1552 & 1555 & 21$\pm$28 & 13 & 2 &  distorted cycle$^*$\\
11 & 1074 & 1080 & 62$\pm$41 & 9 & 0 &   & 54 & 1565 & 1570 & 9$\pm$26 & 9 & 1 &  \\
12 & 1083 & 1087 & 55$\pm$39 & 10 & 0 &  distorted cycle$^*$ & 55 & 1574 & 1578 & 44$\pm$37 & 10 & 2 &  \\
13 & 1093 & 1099 & 64$\pm$46 & 16 & 0 &  distorted cycle$^*$  & 56 & 1584 & 1591 & 34$\pm$41 & 11 & 4 &  \\
14 & 1109 & 1119 & 54$\pm$34 & 16 & 1 &   & 57 & 1595 & 1601 & 62$\pm$46 & 14 & 4 &  \\
15 & 1125 & 1137 & 115$\pm$59 & 17 & 1 &   & 58 & 1609 & 1615 & 14$\pm$32 & 11 & 4 &  \\
16 & 1142 & 1148 & 46$\pm$45 & 11 & 5 &   & 59 & 1620 & 1626 & 13$\pm$28 & 12 & 4 &  \\
17 & 1153 & 1158 & 14$\pm$28 & 10 & 5 &   & 60 & 1632 & 1638 & 1$\pm$23 & 9 & 0 &  \\
18 & 1163 & 1169 & 49$\pm$42 & 13 & 4 &   &  \textit{61 } & \textit{ 1641 } & \textit{ 1646 } & \textit{  6 $\pm$ 22  } & \textit{ 9 } & \textit{ 0 } & \textit{ Maunder minimum}\\
19 & 1176 & 1184 & 49$\pm$45 & 12 & 3 &   &  \textit{62 } & \textit{ 1650 } & \textit{ 1655 } & \textit{  -1 $\pm$ 22  } & \textit{ 8 } & \textit{ 0 } & \textit{ Maunder minimum}\\
20 & 1188 & 1193 & 106$\pm$59 & 11 & 4 &   &  \textit{63 } & \textit{ 1658 } & \textit{ 1663 } & \textit{  -2 $\pm$ 21  } & \textit{ 10 } & \textit{ 0 } & \textit{ Maunder minimum}\\
21 & 1199 & 1205 & 54$\pm$38 & 11 & 5 &   &  \textit{64 } & \textit{ 1668 } & \textit{ 1673 } & \textit{  -6 $\pm$ 21  } & \textit{ 10 } & \textit{ 0 } & \textit{ Maunder minimum}\\
22 & 1210 & 1214 & 28$\pm$36 & 9 & 3 &   &  \textit{65 } & \textit{ 1678 } & \textit{ 1682 } & \textit{  13 $\pm$ 23  } & \textit{ 8 } & \textit{ 0 } & \textit{ Maunder minimum}\\
23 & 1219 & 1225 & 10$\pm$26 & 13 & 2 &   &  \textit{66 } & \textit{ 1686 } & \textit{ 1691 } & \textit{  -7 $\pm$ 24  } & \textit{ 10 } & \textit{ 0 } & \textit{ Maunder minimum}\\
24 & 1232 & 1237 & 68$\pm$55 & 9 & 4 &   &  \textit{67 } & \textit{ 1696 } & \textit{ 1701 } & \textit{  -4 $\pm$ 21  } & \textit{ 13 } & \textit{ 0 } & \textit{ Maunder minimum}\\
25 & 1241 & 1246 & 58$\pm$47 & 10 & 4 &   &  \textit{68 } & \textit{ 1709 } & \textit{ 1715 } & \textit{  -9 $\pm$ 19  } & \textit{ 8 } & \textit{ 0 } & \textit{ Maunder minimum}\\
26 & 1251 & 1256 & 46$\pm$49 & 11 & 3 &   &  {69 } & { 1717 } & { 1723 } & {  35 $\pm$ 36  } & { 16 } & { 0 } & distorted cycle$^*$\\
27 & 1262 & 1271 & 24$\pm$39 & 13 & 1 &   & 70 & 1733 & 1739 & 71$\pm$47 & 12 & 4 &  \\
28 & 1275 & 1279 & 20$\pm$40 & 12 & 2 &  1279 AD event  & 71 & 1745 & 1749 & 6$\pm$25 & 9 & 1 &  \\
29 & 1287 & 1291 & 17$\pm$39 & 10 & 3 &   & 72 & 1754 & 1759 & 64$\pm$48 & 9 & 4 &  SC1$^\dagger$\\
\textit{30 } & \textit{ 1297 } & \textit{ 1301 } & \textit{  -5  $\pm$ 22  } & \textit{ 10 } & \textit{ 0 } & \textit{ Wolf minimum}  & 73 & 1763 & 1766 & 51$\pm$43 & 9 & 2 &  SC2$^\dagger$\\
\textit{31 } & \textit{ 1307 } & \textit{ 1313 } & \textit{  7 $\pm$ 21  } & \textit{ 11 } & \textit{ 0 } & \textit{ Wolf minimum} & 74 & 1772 & 1778 & 134$\pm$54 & 11 & 3 & SC3$^\dagger$\\
\textit{32 } & \textit{ 1318 } & \textit{ 1320 } & \textit{  -7 $\pm$ 20  } & \textit{ 8 } & \textit{ 0 } & \textit{ Wolf minimum} & 75 & 1783 & 1789 & 115$\pm$55 & 12 & 2 & SC4$^\dagger$\\
\textit{33 } & \textit{ 1326 } & \textit{ 1334 } & \textit{  7 $\pm$ 22  } & \textit{ 14 } & \textit{ 0 } & \textit{ Wolf minimum} & 76 & 1795 & 1801 & 14$\pm$27 & 11 & 4 & SC5$^\dagger$\\
34 & 1340 & 1344 & 16$\pm$37 & 11 & 0 &   & 77 & 1806 & 1815 & 5$\pm$28 & 14 & 2 & SC6$^\dagger$\\
35 & 1351 & 1357 & 108$\pm$77 & 12 & 3 &   & 78 & 1820 & 1825 & 37$\pm$44 & 9 & 4 & SC7$^\dagger$\\
36 & 1363 & 1369 & 114$\pm$65 & 12 & 3 &   & 79 & 1829 & 1833 & 114$\pm$57 & 10 & 4 & SC8$^\dagger$\\
37 & 1375 & 1381 & 35$\pm$39 & 11 & 4 &   & 80 & 1839 & 1844 & 54$\pm$40 & 8 & 2 & SC9$^\dagger$\\
38 & 1386 & 1391 & -4$\pm$21 & 9 & 2 &   & 81 & 1847 & 1852 & 72$\pm$47 & 9 & 4 & SC10$^\dagger$\\
39 & 1395 & 1400 & 8$\pm$24 & 10 & 2 &   & 82 & 1856 & 1861 & 91$\pm$50 & 9 & 4 &  SC11$^\dagger$\\
\textit{40 } & \textit{ 1405 } & \textit{ 1418 } & \textit{  7  $\pm$ 23  } & \textit{ 17 } & \textit{ 0 } & \textit{Sp\"orer minimum } & 83 & 1865 & 1872 & 73$\pm$42 & 13 & 3 &  SC12$^\dagger$\\
\textit{41 } & \textit{ 1422 } & \textit{ 1435 } & \textit{  -3 $\pm$ 24  } & \textit{ 10 } & \textit{ 0 } & \textit{Sp\"orer minimum } & 84 & 1878 & 1883 & 24$\pm$29 & 10 & 3 & SC13$^\dagger$\\
\textit{42 } & \textit{ 1432 } & \textit{ 1436 } & \textit{  -5 $\pm$ 26  } & \textit{ 10 } & \textit{ 0 } & \textit{Sp\"orer minimum} & 85 & 1888 & 1894 & 23$\pm$32 & 12 & 4 & SC14$^\dagger$\\
\textit{43 } & \textit{ 1442 } & \textit{ 1453 } & \textit{  -9 $\pm$ 25  } & \textit{ 15 } & \textit{ 0 } & \textit{Sp\"orer minimum}  &  \\ &  &  & $\pm$ &  &  & \\	
\hline
\end{tabular}
\\$^*$ Transition from grand minimum to normal activity modes.
\\$^\dagger$ Standard Schwabe cycles (www.ngdc.noaa.gov/stp/solar/\-solardataservices.html).
\label{Tab:cycles}
\end{table*}

\subsection{Activity levels}

\subsubsection{Distribution of activity levels}

\begin{figure}
\centerline{\includegraphics[width=\columnwidth]{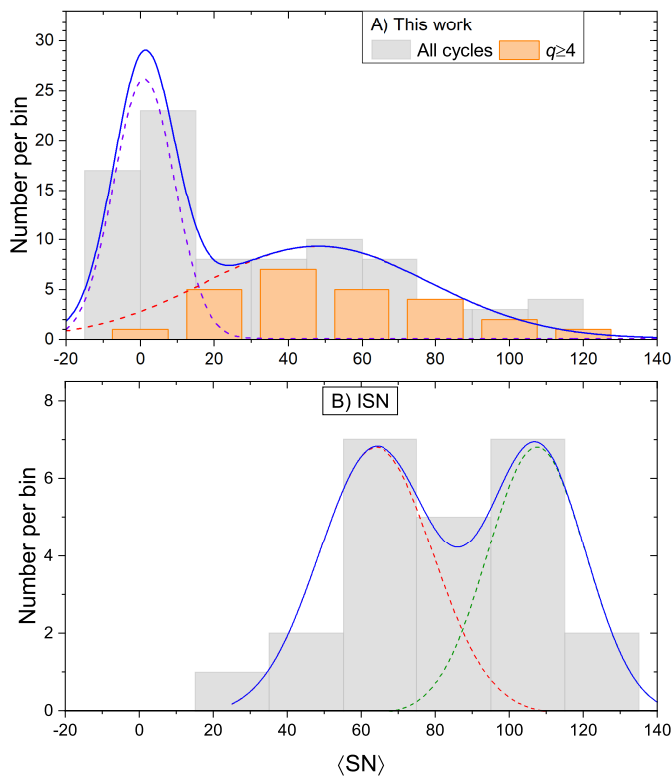}} 	
\caption{Distribution of cycle-averaged $\langle$SN$\rangle$ sunspot numbers along with fitted bimodal Gaussians.
Panel A: Sunspot numbers reconstructed here for 971\,--\,1900 (85 full cycles), the two curves are:
 grand-minimum mode (magenta dashed line, mean $m$=1, $\sigma$=8) and normal mode (red dashed line $m$=48, $\sigma$=31).
Orange bars correspond to the well-defined cycle $q\geq 4$ ($m$=49, $\sigma$=36, 25 cycles).
Panel B: ISN (v.2) sunspot numbers for 1750\,--\,2019 (24 full cycles) and includes two modes: normal one ($m$=64, $\sigma$=16)
 and grand-maximum one ($m$=107, $\sigma$=13).
}
\label{Fig:SN_dist}
\end{figure}

The distribution of the cycle-averaged sunspot numbers is shown in Figure~\ref{Fig:SN_dist}A, where two clearly separated
 modes can be observed: the grand-minimum mode with $\langle{\rm SN}\rangle <20$ (well fitted with a Gaussian with the mean
 $m$=1 and $\sigma$=8), and a normal mode with $m=$48 and $\sigma$=31.
The clear separation of the modes confirms the earlier finding that grand minima form a special mode of solar activity,
 as discovered by \citet{usoskin_AAL_14} using the INTCAL data for the last three millennia.
The distribution of $\langle{\rm SN}\rangle$ for the well-defined ($q$$\geq$4) cycles can be fitted with a single-mode
 Gaussian distribution ($m$=49, $\sigma$=36).

For comparison, the distribution of the cycle-averaged $\langle{\rm SN}\rangle$ for the direct sunspot number series
 (ISN v.2, 1750\,--\,2019) is shown in Figure~\ref{Fig:SN_dist}B.
Although the statistics is low (24 full cycles), two distinct modes can be found: a normal one ($m$=63 and $\sigma$=16),
 and the grand-maximum one ($m$=107, $\sigma$=13).

It is important that the distributions of the normal-mode activity, centered at about 50 $\langle$SN$\rangle$,
 cannot be distinguished in a statistical sense ($p$-value 0.07) for the direct SN observations and the reconstructed one.
For the period of direct overlap between the series (1750\,--\,1900), the mean reconstructed $\langle$SN$\rangle$ value is $62\pm 10$,
 while it is $77\pm 8$ for the ISN (v.2) series, implying that the difference is systematic but insignificant ($p$=0.24).
A grand-minimum mode is clear in SN reconstructed here for the last millennium ($\approx$50\% of time) but absent in the ISN series, where
 only a short and shallow Dalton minimum is present.
On the other hand, the ISN dataset contains the Modern grand maximum in the second half of the 20th century,
 while no grand maxima are found for the period 970\,--\,l900.
A few high cycles can be seen in the reconstructed SN ca. 990, 1200, 1370 and 1790, but the Modern grand maximum
 is unique in the combination of both the level and length, over the last millennium, in agreement with earlier
 findings \citep[e.g.,][]{usoskin_PRL_03,solanki_Nat_04,usoskin_AA_07}.
The very high cycles in the beginning of the series may be related to the not-yet-relaxed carbon-cycle model.

\subsubsection{Grand minima}

The last millennium was not very typical in solar activity as it covers the low phase of the Halstatt cycle with
 five Grand solar minima (Oort, Wolf, Sp\"orer, Maunder and Dalton -- see Figure~\ref{Fig:SN}A) with a total duration
 of about 430 years \citep{usoskin_AA_16,brehm21}.
Thus, the Sun spent nearly half of the last millennium in the grand minimum mode, while the average fraction is about 17\%
 for a 10000-year period \citep{usoskin_AA_07,inceoglu15}.
Out of these 430 years, about 220 years can be identified as deep-minimum phases when the level of activity drops
 very low, below the sunspot formation threshold $F{\rm o}$$<$$2.5\cdot 10^{14}$ Wb: 1300\,--\,1330 (Wolf minimum),
 1410\,--\,1540 (Sp\"orer minimum) and 1650\,--\,1710 (Maunder minimum), as marked in Table~\ref{Tab:cycles}.

The reconstructed mean level of solar activity during the major grand minima is consistent with a SN of zero within the
 error bars (Figure~\ref{Fig:SN}A) for the Wolf, Sp\"orer and Maunder minima, but is of the order of 10 for the shorter
 and shallower Oort and Dalton minima (in agreement with the direct SN data during the Dalton minimum).

\begin{figure}
\centerline{\includegraphics[width=\columnwidth]{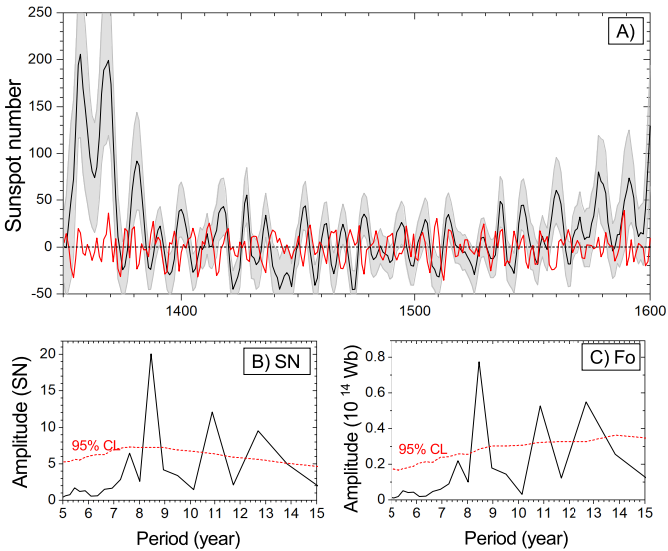}} 	
\caption{Reconstructed solar activity around the Sp\"orer minimum.
Panel A: The black curve with grey-shaded error bars represents the main reconstruction and is identical to that in Figure~\ref{Fig:SN_panels},
 while the red curve depicts a reconstruction based on one realization of the simulated noise series (see text).
Panel B: FFT spectrum (amplitude) of the main SN reconstruction (shown in Panel A) for the period 1400\,---\,1550, while the red dotted line
 depicts the upper 95th percentile of the FFT spectra of 1000 noise-based reconstructions for the same period (see text).
Panel C: The same as panel B but for OSF.
}
\label{Fig:MM_test}
\end{figure}

Shapes of individual cycles are poorly defined (quality flag $q$ is low) during the deep phases of the major minima.
To check the robustness of the cycle assessment we have performed the following test.
We produced 1000 synthetic noise-based series of $Q^{14}$C with the statistical distribution (mean 2 and $\sigma=0.22$, both in units of at/g/cm$^2$)
 corresponding to those \citep{brehm21} of the actual dataset during the Sp\"orer minimum (1400\,--\,1550).
The synthetic series was processed further by applying the same method as described in Section~\ref{Sec:method},
 viz. identically to the main reconstruction.
One such realization of the SN reconstructed from purely noisy series is shown in Figure~\ref{Fig:MM_test}A as the red curve,
 along with the main SN reconstruction for the period of the Sp\"orer minimum.
It exhibits a seemingly oscillating pattern with a typical length of 3\,--\,20 years and an amplitude of 10\,--\,20 in SN,
 some of these oscillations are comparable to the cycles reconstructed from the real data.
For each of these 1000 noise-based reconstructions, we computed the FFT (amplitude) spectrum, and then took the upper 95th percentile
 of them, as shown by the red dotted line in Figure~\ref{Fig:MM_test}B.
One can see that the spectrum is nearly flat, with the amplitude being 5\,--\,7 in SN.
The FFT spectrum of the final reconstructed SN series, based on the real data, for the period of 1400\,--\,1550
 is shown in black and contains peaks around 8.5, 11 and 13 years.
These peaks are significantly higher than the 95\% confidence level.
The peak at about 9 years is consistent with a result for the Maunder minimum \citep{vaquero15} based on sunspot observation.

Therefore, we conclude that, while the presence of the Schwabe cycle during the Sp\"orer minimum is statistically
 significant, individual cycles are not robustly defined.
Accordingly, we mark all cycles inside the deep grand minima as unreliable (quality flag $q$=0) and exclude them from the
 cycle length analysis.
A similar pattern can be observed in the OSF evolution during a grand minimum (Figure~\ref{Fig:MM_test}C).

Cycles outside the grand minima are defined fairly reliably.
However, there are several distorted or merged cycles:
 ca. 1100, 1360, 1560 and 1720 (see Table~\ref{Tab:cycles}), which occurred
 shortly after the transition from a grand minimum to normal activity.
This may be related to the fact that the relation between the heliospheric
 modulation of cosmic rays and solar magnetic activity can be inverted during grand minima, as shown observationally
 for the Maunder minimum \citep{beer98,usoskin_JGR_MM_01} and proposed theoretically \citep{owens12}.

The level of the reconstructed activity is very low, below the sunspot-formation threshold,
 during the Maunder minimum, in agreement with all other datasets \citep{vaquero15,usoskin_MM_15}.
It is interesting that, according to the new reconstruction, the deep phase of the Maunder minimum extended until
 about ca. 1710 in agreement with earlier results \citep{eddy76,usoskin_MM_15,vaquero_NA_15} but in contradiction with
 the formal ISN data series (Figure~\ref{Fig:SN}C).

\subsubsection{Comparison with direct SN series}
\label{Sec:comp}

Comparison of the SN reconstructed here with other series based on direct sunspot observations is shown in Figure~\ref{Fig:SN}.
All series were reduced to the ISN(v.2) definition.
The lower-bound is represented by the GSN \citep{hoyt98}, which yields the lowest sunspot numbers among all series,
 while ISN(v.2) \citep{clette16} forms the upper bound.

Figure~\ref{Fig:SN}A presents a comparison of smoothed (15-yr first SSA component) series.
The reconstructed activity lies between the two bounds for the period of 1750\,--\,1900.
It is closer to the GSN series in the 19th century, while being close to ISN (v.2) in the second half of the 18th century.
ISN (v.2) falls $>1\sigma$ above the reconstructed data around 1830\,--\,1870, ca. 1810 and before 1730,
 while GSN lies $>1\sigma$ below the reconstructed series between ca. 1730 and 1780.
Thus, the reconstructed SN is consistent with both ISN and GSN being somewhat close to the GSN one.
The reconstructed SN goes quite low around 1900, probably related to the somewhat uncertain Suess-effect correction.

Figure~\ref{Fig:SN}C focuses on the last three centuries.
One can see that we correctly reconstruct most of individual solar cycles, at least in the sense of times of their
 maxima and minima, which agree within $\pm$2 years with those in the actual sunspot numbers.
Exceptions are the periods ca. 1840, when an extra cycle appeared distorting the neighbouring cycles,
 ca. 1725 when a long, possibly merged cycle appeared, and ca. 1770 when a cycle is distorted by a sudden in drop.
The former two distortions may be related to a transition between grand-minimum (Dalton and Maunder minima, respectively) and normal modes of activity,
 as discussed above, while the latter one is related to a strong jump in $^{14}$C production.
Thus, out of eighteen cycles during the period 1700\,--\,1900, fifteen are reproduced correctly, three are distorted, but the
 total number of cycles is preserved.

Magnitudes of the cycles vary: while some reconstructed cycles cover the full extent of the directly measured sunspot cycles which
 drop to close to zero near activity minima, see ca. 1860, some have small magnitude, such as ca. 1790 (consequently the reconstructed
 series tends to underestimate the true amplitude of the SN series at such times).
However, as visible from Figure~\ref{Fig:SN}A, the mean level is preserved.

The SN series obtained here does not follow any single SN series in the literature,
 but shows elements of different ones, being closer to one at some times, but to another at other times.
The mean difference between the cycle-averaged SN values reconstructed here and those based on direct solar
 observations (all series are reduced to the ISN v.2 definition) for the 19th century are:
 15$\pm$3.2 for ISN (v.2) implying a significant systematic difference (ISN v.2 is systematically higher),
 -4.8$\pm$3.3 for GSN implying that it is systematically lower,
 and 2.9$\pm$3.8 for the series by \citet{chatzistergos17} implying their mutual consistency.

We conclude that the method correctly reconstructs the mean level and minimum/maximum dates of individual cycles,
 but the exact magnitudes of some sunspot cycles may be distorted.

\subsection{Cycle lengths}
\label{Sec:length}

\begin{figure}
\centerline{\includegraphics[width=\columnwidth]{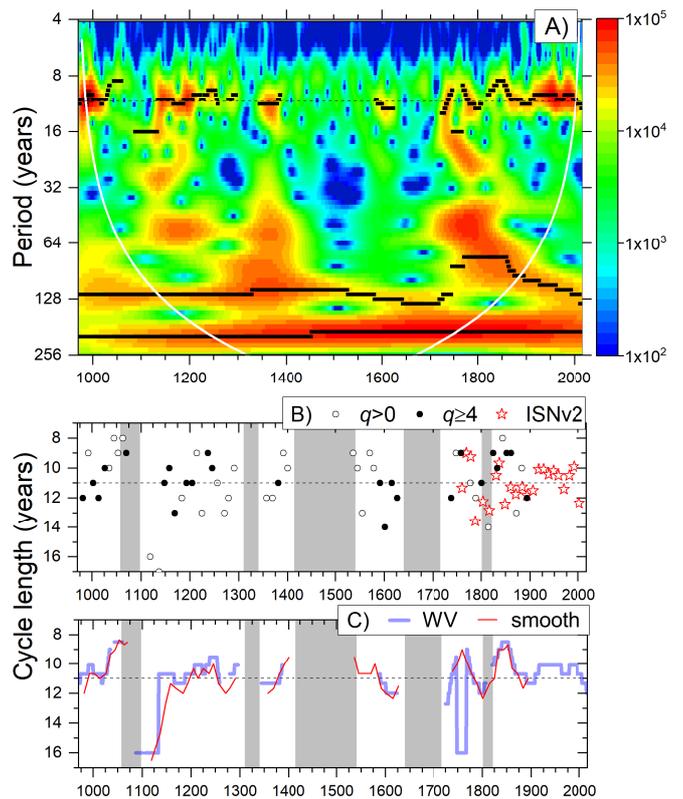}} 	
\caption{Panel A: Wavelet power spectrum (Morlet basis, $k=6$) of the reconstructed SN series, extended by ISN (v.2) after 1900.
The white curve bounds the cone of influence (COI) where the spectrum is unreliable because of the
 proximity to the edges of the series.
Black dots denote local maxima of the power ($>$10$^4$) in the period bands 7\,--\,16, 75\,--\,140 and 180\,--\,250 years.
The horizontal dashed line marks the location of the 11-year period.
Panel B: Cycle length (minimum-to-minimum) as function of time (assigned to the cycle maximum year) for resolvable ($q$$>$0, 56 cycles)
 reconstructed cycles (open dots) and for well-defined cycles ($q$$\geq$4, 25 cycles, filled circles).
 Red stars are the cycle lengths, rounded to the nearest integer (to be reduced to the annual resolution),
  in the ISN(v.2) series (http://www.sidc.be/silso/cyclesminmax) for 1750\,--\,2019.
 The dashed line represents the mean cycle length of 11 years over the ISN series.
 The grey shading denotes grand minima of solar activity.
Panel C: Comparison between the smoothed cycle-length evolution.
  The red \textit{smooth} curve is the 3-point running mean of the individual cycle lengths ($q$$>$0).
  The light blue \textit{WV} curve is the wavelet-defined period (identical to the upper black curve in panel A).
}
\label{Fig:SN_WV}
\end{figure}

Wavelet power spectrum of the reconstructed sunspot activity is shown in Figure~\ref{Fig:SN_WV}A with `ridges'
 (local maxima of the spectrum for each year) shown as black dots.
The reconstructed series has been extended to cover the 20th century with the ISN(v.2) series.

One can see the dominant $\approx$11-year Schwabe quasi-periodicity visible as the yellow-red ribbon at 8\,--\,16 years with several
 breaks corresponding to grand minima (Oort, Wolf, Sp\"orer and Maunder minima) when the reconstructed cycles
 have low amplitude and are not reliable.
It was proposed earlier that the typical length of the Schwabe cycle tends to increase
 before/during grand minima \citep[e.g.,][]{fligge99,miyahara_JGR_06}, as based on poorer statistics.
We confirm this pattern for the Maunder and Dalton minima but not for other minima
 (see Figure~\ref{Fig:SN_WV}).

Another pronounced periodicity corresponds to the Suess/de Vries cycle of about 210 years,
 which appears very stable without any significant fluctuation.
The 210-year periodicity manifests itself mostly as recurrence of grand minima within a
 cluster \citep{usoskin_AA_07,usoskin_AA_16}.

The so-called Gleissberg or centennial cycle is visible as a broad pattern at about 120 years,
 extending to shorter periods during 1700\,--\,1800, and being not very stable, in agreement with earlier
 findings \citep[e.g.,][]{ogurtsov02}.

A shortcoming of the wavelet analysis is that it yields an estimate of the general variability of the periods and
 does not provide information on individual cycles.
Using dates of the solar cycle minima (Table~\ref{Tab:cycles}), one can analyse
 lengths of individual solar cycles as shown in Figure~\ref{Fig:SN_WV}B for all resolvable cycles ($q$$>$0, open circles),
 well-defined cycles ($q$$\geq$4, filled circles) and the directly observed cycles (red stars).
The length of individual resolvable cycles ($q$$>$0) generally agrees
 (red vs. light blue curves in Figure~\ref{Fig:SN_WV}C)
 with the wavelet-based definition after a 3-point smoothing of the former,
 that roughly corresponds to the wavelet package extension.
This confirms that our minimum-to-minimum cycle length definition is robust, since the wavelet-based definition considers
 the full variability within the wavelet package, not only minima or maxima.

\begin{figure}
\centerline{\includegraphics[width=\columnwidth]{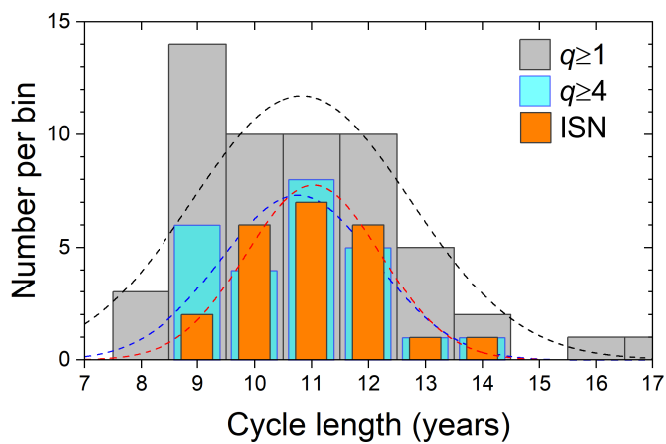}} 	
\caption{Distribution of cycle lengths for all resolvable cycles with $q$$>$0 (grey bars),
 well-defined cycles ($q$$\geq$4, blue) and ISN (v.2) series (http://www.sidc.be/silso/cyclesminmax) for 1750\,--\,2019.
 Dashed lines represent the fitted normal distributions.
}
\label{Fig:C_L_dist}
\end{figure}

The distribution of the cycle lengths is shown in Figure~\ref{Fig:C_L_dist} along with fitted normal distributions:
 $10.8\pm 1.9$ years for 56 resolvable cycles with $q$$>$0, (grey bars), $10.8\pm 1.4$ years for well-defined cycles ($q$$\geq$4, 25 cycles, blue bars),
 and $11.0\pm 1.1$ years for the ISN (23 cycles, orange bars).
We have tested the hypothesis of the equality of the cycle length distributions using the $z$-score test.
All the three distributions (outside of the grand minima) cannot be considered as different at any reasonable significance level.
Thus, we conclude that the cycle-length distribution of the reconstructed cycles outside deep grand minima
 is statistically consistent  with that for the directly observed solar cycles after 1750.

\subsection{Waldmeier rule}

The Waldmeier rule states that cycles with faster rising SN are stronger.
This rule is built on a statistically significant correlation
 between the length of the rising phase and the peak height of the cycles \citep{hathawayLR,usoskin_Wald_21}.
Since the amplitude of the cycle SN$_{\rm max}$ is not well determined here, we considered a more robust quantity of
 the cycle-averaged sunspot number $\langle{\rm SN}\rangle$ from Table~\ref{Tab:cycles}.
The relation between the length of the scending phase $T_{\rm as} =Y_{\rm max}-Y_{\rm min}$ (in years) and $\langle{\rm SN}\rangle$
 for all 85 cycles appears insignificant (Pearson linear correlation coefficient $r$$\approx$0).
However, when only well-defined cycles (quality flag $q\geq 4$) are considered, the Waldmeier rule appears
 highly significant ($r$=-0.58$^{+0.12}_{-0.16}$, $p$-value=0.001, $N$=25) as

$\langle{\rm SN}\rangle =(-26\pm 16)\cdot T_{\rm as}+(197\pm90)$,

\noindent
The Waldmeier rule for the direct SN series defined in the same way (i.e., the cycle-mean SN vs. length of the ascending phase
 rounded to an integer) yields the following relations

$\langle{\rm SN}\rangle = (-14.5\pm 7)\cdot T_{\rm as}+(150\pm 35)$

\noindent ($r=-0.67\pm0.15,\,\, p<0.001$) for ISN (v.2) and

$\langle{\rm SN}\rangle = (-12.8\pm 9)\cdot T_{\rm as}+(137\pm 45)$

\noindent ($r=-0.53\pm0.15$, $p=0.005$) for GSN.

Thus, we confirm that the Waldmeier rule is valid also on the millennial scale, at least in the sense
 of the cycle-average SN rather than cycle-peak SN, which is poorly defined in the reconstructed SN timeseries.

\section{Summary and Conclusions}

A new quantitative reconstruction of annually resolved solar activity, in the form of sunspot numbers
 (at least outside grand minima) with full uncertainty
 assessment, is presented for the period 971\,--\,1900.
For the first time, individual solar cycles are presented for the whole of the last millennium, more than doubling the
 existing statistics of solar cycles.
Overall, 85 solar cycles are reconstructed (the mean length 10.8$\pm$1.9 years), of which 25 cycles
 are well-resolved (the mean cycle length 10.8$\pm$1.4 years), 10 cycles are reasonably defined,
 21 are poorly defined, and 29 cycles cannot be reliably identified.
The unresolvable cycles correspond to the deep-minimum phase of solar activity
 during grand minima.
The periods of low activity were abnormally frequent (about 40\% of the time)
 during the last millennium covering a cluster of four grand minima \citep{usoskin_AA_16}.
The new reconstruction agrees well, within the uncertainties, with the estimates of the sunspot numbers based
 on direct telescopic observations during the 18th and 19th centuries, in both the mean level and the mean cycle length.
This greatly increases the number of the known solar cycles, from 36 cycles known for the period between 1610\,--\,2019, including
 24 well-defined ones after 1755, 8 poorly resolved and 4 unresolvable cycles around the Maunder minimum,
 to 96 cycles (971\,--\,2019) including 50 well- and reasonably defined cycles, 17 poorly defined cycles,
 and 29 individually unresolvable cycles for the last millennium.

The grand minima form a separate mode of solar activity with solar activity dropping below the sunspot formation threshold
 during their deepest phases.
The deepest grand minima were the Sp\"orer minimum with the deep phase covering 1410\,--\,1540,
 Maunder minimum (1650\,--\,1710) and Wolf minimum (1300\,--\,1330), while the shorter Oort and Dalton minima
 did not drop to such a deep level \citep[cf.][]{brehm21}.
A cyclic variation of solar activity was found even during the deep phases of the grand minima,
 agreeing with and extending earlier results \citep{beer98,miyahara06,miyake_JGR_13}, but their significance
 is low since their magnitude is less than $1\sigma$ uncertainty.
However, the variation in solar activity at such times was at a level that it produced none or at the most few sunspots.

The new data confirms that the Maunder minimum extended until at least 1710 in agreement with other datasets but
 in contradiction to the ISN series.
It is found that in the record constructed from $^{14}$C data distorted cycles appear at the transition phase between grand minima and normal activity,
 allowing for the possibility that the relation between solar activity and heliospheric modulation of cosmic rays
 may be different for the two modes \citep[e.g.,][]{owens12}.
Alternatively, at the end of a grand minimum the solar dynamo may behave in ways not treated properly
 by the model used here to reconstruct SN.

Three sudden increases of $^{14}$C production, in 994, 1052 and 1279 AD were removed from the initial dataset before
 the reconstruction.
The fact that this step restores nearly perfectly cyclic solar variability around these dates supports our method of treating these events.
The event of 994 AD, the second largest SPE known to date, took place at the early declining phase of a moderately strong
 solar cycle; the event of 1052 AD corresponded to the maximum phase of a moderate solar cycle; the event of 1279 AD took place
 at the maximum phase of a moderate solar cycle, but the phase may be affected by the removal procedure.

The validity of the empirical Waldmeier rule (cycles with faster ascending phase tend to be stronger, viz. have greater mean SN)
 is confirmed at a significant statistical level for well-defined cycles (quality flag $q$$\geq$4) but is blurred
 out when all cycles are considered.

The new, first quantitative sunspot-number reconstruction at the annual time scale with full uncertainties,
 building on the important work of \citet{brehm21} and making use of a significantly improved reconstruction technique,
 opens up new avenues  in solar and solar-terrestrial studies with implications for the solar dynamo (specifically the
 transition between grand-minimum and normal activity modes), reconstructions of solar irradiance, etc.

The results presented here form a big step forward compared with earlier reconstructions of solar activity,  which,
 with few exceptions \citep{stuiver93,miyahara04,fogtmann17}, have only provided decadal resolution over the last millennium or longer
 \citep[e.g.][]{usoskin_PRL_03,solanki_Nat_04,vonmoos06,steinhilber12,wu18}.
The record of individually resolved solar cycles has been nearly tripled (doubled for well-resolved cycles) providing a basis for more precise
 solar and solar-terrestrial studies extending now over the whole last millennium.

\acknowledgement{
This work was partly supported by the Academy of Finland (Projects ESPERA no. 321882).
}


%
%
\bibliographystyle{aa}


\end{document}